\documentclass[amsfonts, amssymb, amsmath, reprint, showkeys, nofootinbib, twoside]{revtex4-1}

\usepackage[english]{babel}
\usepackage[utf8]{inputenc}
\usepackage{booktabs}

\usepackage{float}
\makeatletter
\let\newfloat\newfloat@ltx
\makeatother

\usepackage{algorithm}
\usepackage{algpseudocode}

\usepackage{caption}
\usepackage[figurename=FIG.]{caption}

\usepackage[colorinlistoftodos, color=green!40, prependcaption]{todonotes}

\usepackage[pdftex, pdftitle={Article}, pdfauthor={Author}]{hyperref} 
\begin{document}
\title{Self Organized Critical Dynamic on the Sierpinski Carpet}

\author{Viviana Gómez}
\email{v.gomez6@uniandes.edu.co}

\author{Gabriel Téllez}
\email{gtellez@uniandes.edu.co}

\affiliation{Physics Department, Universidad de los Andes, Bogotá, Colombia.}

\begin{abstract}

Self-organized criticality is a dynamical system property where, without external tuning, a system naturally evolves towards its critical state, characterized by scale-invariant patterns and power-law distributions. In this paper, we explored a self-organized critical dynamic on the Sierpinski carpet lattice, a scale-invariant structure whose dimension is defined as a power-law with a non-integer exponent, i.e. a fractal. To achieve this, we proposed an Ising-BCP (bond-correlated percolation) model  as the foundation for investigating critical dynamics. Within this framework, we outlined a feedback mechanism for critical self-organization and followed an algorithm for its numerical implementation. The results obtained from the algorithm demonstrated enhanced efficiency when driving the Sierpinski carpet towards critical self-organization compared to a two-dimensional lattice. This efficiency was attributed to the iterative construction of the lattice and the distribution of spins within it. The key outcome of our findings is a novel dependence of self-organized criticality on topology for this particular model, which may have several applications in fields regarding information transmission.

\end{abstract}


\maketitle

\section{Introduction}

The concept of self-organized criticality (SOC) describes how complex systems can spontaneously organize themselves into a critical state. This means that the critical state arises from the collective behavior of many interacting components in the system rather than being externally imposed \cite{PhysRevA.38.364}. In recent decades, numerous systems exhibiting SOC have been identified, ranging from simple sandpile models to more complex systems such as neural networks, earthquakes, forest fires, and financial markets \cite{Pu2013Developing,  Sornette1989Self-Organized, Ricotta1999The}. One of the remarkable features of systems undergoing SOC is the emergence of scale-invariant behavior, manifesting patterns that resemble fractal-like structures.\\

The underlying mechanism of SOC is the alternation between a slow-driving process and fast dynamics with dissipation. This creates a feedback loop that guides the system to the critical point of a continuous phase transition \cite{10.3389/fphy.2020.00333}. Percolation models and the Ising model are two classic examples of physical and mathematical systems that exhibit continuous phase transitions. Several previous studies have explored the conversion of the second-order phase transitions in these two models into SOC \cite{Sornette1992CriticalPT, pruessner2006self, vespignani1997order, PhysRevE.57.5095, dickman2000paths}. The main idea involves establishing a feedback loop that adjusts the control parameter based on the order parameter. In this context, the networks on which these two models are built are called lattices and can have specific geometries. For example, square-lattices, triangle-lattices, or fractal lattices \cite{PhysRevB.58.14387, PhysRevLett.51.2347}. Regarding the latter, examples of fractal structures are the Sierpinski carpet (SC) and the Sierpinski triangle. Both of these fractals are based on an iterative process that involves dividing and removing smaller geometric shapes, creating a self-similar figure characterized by a remarkable degree of order.\\

As mentioned earlier, self-organized critical systems often exhibit a fractal behavior in their activity patterns and in the distribution of critical event sizes. Consequently, it makes sense to think that the topology of a fractal lattice could influence the dynamic of the modeled system, potentially giving rise to intriguing properties such as increased robustness and enhanced functionality. These properties have potential applications in various practical fields, including neural network modeling, transportation network analysis, infectious disease spread prediction, among others \cite{bieberich2002recurrent, guida2007topology, zhang2008fractal}. Nonetheless, to date, the self-organized critical dynamics within models constructed on fractal networks have not been thoroughly examined.\\

In this paper, we delved into the dynamic of SOC in an Ising-BCP model constructed on a Sierpinski fractal lattice, the Sierpinski carpet, and investigated how the fractal topology influenced the system behavior. The primary aim was to comprehend the emergent properties of critical self-organization in fractal systems with hierarchical construction, in contrast to uniform or random systems. To achieve this, a computational algorithm was developed, incorporating a feedback loop that dynamically adjusts the control parameter based on the system's order parameter. The algorithm's effectiveness was evaluated across three different lattice configurations, each with an identical number of spins: the Sierpinski carpet lattice, a square lattice, and a non-self-similar lattice generated randomly. To complement the above, a numerical characterization of the Sierpinski carpet phase transition was carried out. Using finite-size scaling (FSS), as well as the Ising model and percolation model formalism, the fractal dimension of the infinite cluster $d_f$ and the $\tau$ critical exponent, which characterizes the distribution of cluster sizes at the critical threshold,  were calculated.  \\

The structure of this article is as follows: Initially, we introduced the foundational concepts, drawing on \cite{Sornette1992CriticalPT, zapperi1995self, stauffer1979scaling} to outline a mechanism for driving certain systems to self-organize critically via an Ising-percolation mapping. Following this, we delved into the study of phase transitions and critical exponents of the Ising model on the Sierpinski carpet, providing a comprehensive review of the previously published results. We then described the algorithm we implemented, detailing its operational mechanics and performance. In concluding, we engaged in a thorough discussion of the physical implications and practical relevance of our findings, shedding light on the broader impact of our research.

\section{SOC dynamics in an Ising-BCP model}
\label{SOCdynamics}

The most straightforward method to implement a SOC feedback mechanism in the Ising model involves using cooling and heating rates that automatically adjust based on the magnetization value until reaching the critical temperature \cite{PhysRevLett.78.4793}. Nevertheless, according to Pruessner \textit{et al.} \cite{pruessner2006self}, in order to reach a critical state with universal scaling, there must be a relationship between the magnetization and the correlation length, and the cooling and heating rates. Then, the objective is to devise a dynamic process in which the order parameter influences the control parameter based also on the value of the correlation length.\\

\begin{figure}[H]
\centering
\includegraphics[scale=0.4]{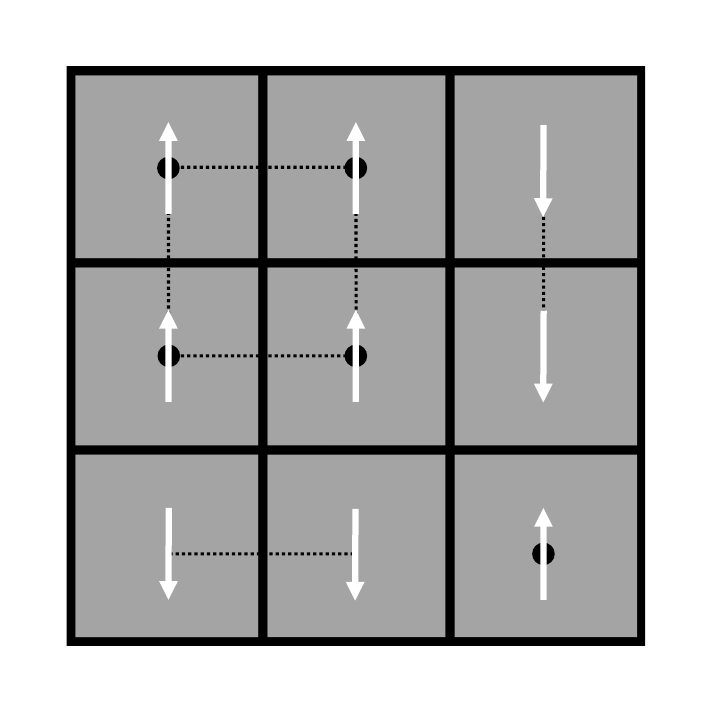}
\caption{Ising Model with a spin connectivity definition that equates its critical point with the emergence of an infinite cluster. Each square represents a spin. If it has a circle inside, it is an spin up or, in the percolation analogy, an occupied site. If it does not have a circle, it is a spin down or unoccupied site. This configuration has one cluster of four spins, two clusters of two spins and one cluster of a single spin.}
\label{fig:conectivity}
\end{figure}

In \cite{Coniglio1980ClustersAI}, Coniglio and Klein proposed a method to map the percolation model onto the Ising model in a manner that aligns the critical point of the Ising model with the one of the percolation model. They noted that a straightforward mapping, where an up spin corresponds to an occupied site and a down spin to an unoccupied site, is generally ineffective because the formation of an infinite cluster does not coincide with the critical temperature of the Ising model. However, they introduced a modification that achieves this alignment (see Figure \ref{fig:conectivity}). This adjustment involves establishing bonds only between nearest-neighbor sites sharing the same state, with a probability $p_B$ determined by
\begin{equation}
    p_B= 1- \exp \left({\frac{-2J}{k_B T}}\right)
    \label{eq:probability}
\end{equation}
where  $J$ is the coupling factor, $T$ is the temperature and $k_B$ is the Boltzmann constant. For simplicity, from now on, all temperatures are expressed in dimensionless units of $[J/k_B]$. 

The Ising model can thus be considered a bond-correlated percolation model (BCPM), and it has been shown that both approaches yield identical critical exponents \cite{hu1984percolation}. This equivalence opens up the possibility of understanding the Ising phase transition in terms of clusters behavior and will be the key to construct a SOC dynamics.\\

Following the approach proposed by Sornette \cite{Sornette1992CriticalPT}, one can create a scenario in which all spins are gradually cooled down at a slow rate, regardless of the system's current state. Simultaneously, each spin can only be heated if there exists a continuous path through the system capable of transferring heat to it. For this path to sustain, it must extend to the edges of the grid to connect with the heat source. The paths are established according to the Coniglio and Klein bonds. This heating condition, dependent on the presence of a connected path, ensures that the system's dynamics converge towards a state where the correlation length becomes infinitely large.\\

In this dynamic, the slow-driving addition of energy into the system, characteristic of SOC behavior, is represented by the heat released through the conducting path, resulting in an increase in temperature. This gradual introduction of energy occurs over a large time scale as the Ising model is allowed to thermalize at this new temperature. Concurrently, the fast relaxation process, where energy is released, corresponds to the slow rate at which each spin is cooled down. The flip of a spin can be thought of as equivalent to the addition of a grain in the sandpile model \cite{PhysRevA.38.364}. At higher temperatures, when more energy enters the system, there is greater movement of spins. An avalanche can then be defined as the readjustment of all spins when the system is perturbed by the motion of a spin, just as the addition of a grain of sand can trigger an avalanche in the sandpile model.\\

To illustrate the above, imagine an Ising model at a temperature $T$ higher than $T_c$. In this case, only finite clusters exist, so there is no heat flow in the system. This means that only the cooling system is active, causing the system to reach $T_c$. Now, imagine an Ising model at a temperature lower than $T_c$. According to the definition of connectivity, there must be a path connecting a fraction of spins to the heat source, causing the temperature to start increasing. As it approaches $T_c$, the connected path will become less dense until it disappears when the correlation length is infinite. This will cause the path to break, and the system will cool down again.\\

In this way, the mapping between the Ising model and the BCP model allows for the creation of a SOC feedback mechanism that adjusts the temperature according to the magnetization and the correlation length. As we will see later, this is straightforward to implement numerically.\\

\subsection{Critical exponents}

The mapping between the Ising model and the BCP model unveils new relationships among the emergent critical exponents. To elucidate this, it is pertinent to remember that in  our Ising-BCP model, each bond is randomly occupied with a probability which depends on the temperature (see equation (\ref{eq:probability})). At the critical probability $p_c$, corresponding to the Ising critical temperature $T_c$, the percolation threshold is reached. This is the point where a vast or infinite cluster first spans the lattice. Below $p_c$, clusters tend to be small and disconnected. Above it, a dominant large cluster overtakes the system, resulting in a more uniform structure. At $p_c$, however, the cluster structure is complex and highly irregular. Indeed, it displays a complex, self-similar structure over several scales. The mass of the infinite cluster (number of occupied sites) scales with its linear size as
\begin{align}
    M_{*}(L) \propto L^{d_f}
    \label{eq:fractaldim}
\end{align}
where $d_f$ is known as its fractal dimension.\\

The fractal dimension is related to some other critical exponents. To introduce these relations, using the percolation formalism, one can define the average density of points connected to one cluster as 
\begin{equation}
    P= \frac{M(L)}{L^d} 
\end{equation}
where $d$ is the dimension of the lattice. Then, the mass of the largest cluster, $M_{*}$, will be 
\begin{equation}
    M_{*}(L) \propto PL^d.
\end{equation}
Near the critical point, the average density of points will follow a power law given by 
\begin{equation}
    P \propto (p-p_c)^{\beta}.
\end{equation}
In this scenario, $P$ corresponds to the order parameter, equivalent to the magnetization in the Ising formalism. Consequently, the exponent $\beta$ remains identical in both formulations. If $L$ is of the same order than the correlation length, $\xi$, it follows that 
\begin{equation}
    PL^d \propto L^{d_f}.
\end{equation}
Finally, using the fact that 
\begin{equation}
    \xi \propto (p - p_c)^{-\nu},
\end{equation}
one gets
\begin{equation}
    d_f = D - \frac{\beta}{\nu}.
    \label{eq:fractal_beta}
\end{equation}

Equation (\ref{eq:fractal_beta}) establishes a connection between the fractal dimension $d_f$ of the largest cluster and the Ising critical exponents $\beta$ and $\nu$ within a finite lattice of linear dimension $L$ {\cite{munoz1999avalanche, suzuki1983phase, coniglio1983droplet}.\\

On the other hand, the critical percolation threshold can also be characterized by taking into account the distribution of clusters sizes. Near the critical point, the clusters size extend up to a maximum cluster size denoted as $s_{max}$. The distribution of these cluster sizes is modulated by a decaying function expressed as
\begin{align}
    n_s \propto s^{-\tau} f(s/s_{max}).
    \label{eq:tau}
\end{align}
where $n_s$ is the number of clusters with size $s$. Moreover, 
\begin{align}
    s_{max} \propto \mid p - p_c \mid^{-1/\sigma}
\end{align}
where $\tau$ and $\sigma$ are critical exponents. According to \cite{stauffer2018introduction, munoz1999avalanche}, the following relations are satisfied
\begin{equation}
    \frac{1}{d_f}= \sigma \nu
\end{equation}
and 
\begin{equation}
    D= \frac{\tau - 1}{\sigma \nu}.
\end{equation}
Therefore,
\begin{equation}
    \tau= \frac{D}{d_f} + 1.
    \label{eq:tau_df}
\end{equation}

\section{Ising model on the Sierpinski carpet}

\subsection{The Sierpisnki carpet}

The Sierpinski carpet (see Figure \ref{fig:SCconstruction}) is a classic example of a fractal and is constructed using an iterative process of eliminating repetitive patterns. For Sierpinski fractals, one can define the generating cell as $SP_s(l^d, N_{\text{occ}}, k)$, where $d$ is the dimension of the initial hypercube (in this context, a square), $l^d$ is the number of hypercubes the initial one is divided into, $N_{\text{occ}}$ is the number of hypercubes preserved post-elimination and $k$ is the number of iterations of this iterative process. The true fractal is obtained when $k \to \infty$, that is, in the thermodynamic limit \cite{monceau2004direct}. The Haussdorf dimension is given by $D_f= \log(N_{\text{occ}})/ \log(l)$. In the case of the SC, $D_f= \log(8)/\log(3)$. 
\begin{figure}[H]
\centering
\includegraphics[scale=0.25]{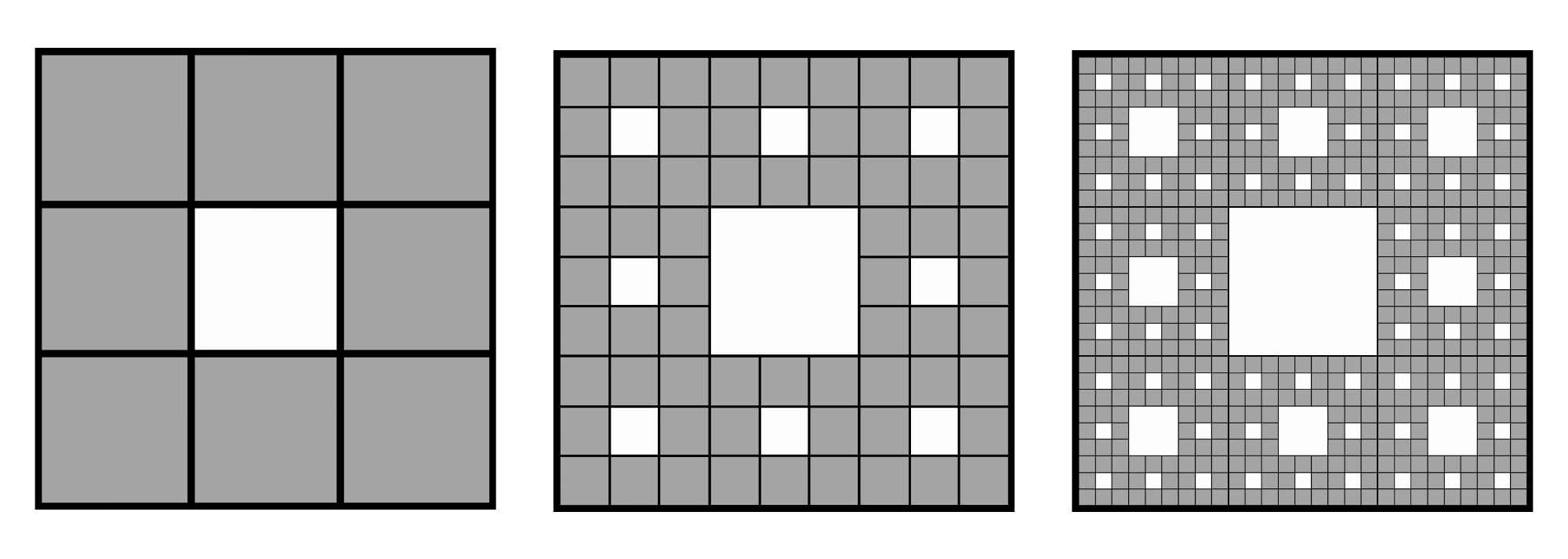}
\caption{Construction of the Sierpinski carpet for $k=3$. }
\label{fig:SCconstruction}
\end{figure}

To isolate an arbitrarily large bounded set of points connected to an arbitrary point in the Sierpinski carpet, the number of significant bonds that must be cut grows as a power of the size of this bounded set. Consequently, the SC exhibits infinite ramification. Previous studies have shown that the Ising model on fractals with infinite ramification displays a second-order phase transition \cite{gefen1980critical}. Moreover, for the case of the SC, this has been confirmed through numerical methods.

\subsection{SC Critical exponents}

Over the last few years, several studies have been conducted to investigate the critical properties of the Ising model on the SC using Monte Carlo simulations. Most of these studies have employed the finite-size scaling (FSS) formalism to calculate a value for the critical temperature and critical exponents \cite{PhysRevB.58.14387, monceau1998magnetic}. However, there is ongoing debate about the effectiveness of this method, with some authors asserting that each stage of segmentation signifies an autonomous thermodynamic system. This challenges the notion of treating it as a scaled iteration of the previous one, which is the fundamental idea behind FSS. In response to the earlier discussion, alternative methods have been developed for calculating critical parameters. Among these are the slope method \cite{pruessner2001monte} and the short-time dynamics (STD) method \cite{bab2005critical}. One approach used to assess the accuracy of simulations involves the Ising model hyperscaling law. Once the critical exponents are determined, it becomes possible to estimate the effective dimension of the system using the formula
\begin{align}
    D_{\text{eff}}= 2 \frac{\beta}{\nu} + \frac{\gamma}{\nu}
    \label{eq:hyperscaling1}
\end{align}
where  $\nu$ is the critical exponent associated with the correlation length, $\beta$ is the critical exponent associated with the magnetization and $\gamma$ is the critical exponent associated with the magnetic susceptibility. Different simulations have provided evidence that this law holds in fractal dimensions. Bab \textit{et al.} \cite{bab2005critical} and Monceau \textit{et al.} \cite{monceau1998magnetic} provided a summary of the most significant results according to different methods. Table \ref{tab:criticalexponents} compiles the results most pertinent to our study.\\

\begin{table}[H]
\centering
\begin{tabular}{@{}|c|c|c|c|c|c|c|c|c|@{}}
\hline
Lattice & $T_c$ & $\nu$ & $\beta$ & $\gamma$ & $D_{\text{eff}}$ & BC & Method & Reference \\ \hline
SC      & 1.481 & 1.70  & 0.136   & 2.94     & 1.89    & P  & FSS    & \cite{PhysRevB.58.14387}   \\ \hline
SC      & 1.482 & 1.73  & 0.254   & 2.81     & 1.92    & F  & FSS    & \cite{monceau1998magnetic} \\ \hline
SC    & 1.494 & 1.39  & 0.121   & 2.22     & 1.77    & P  & STD    & \cite{bab2005critical}     \\ \hline
Square     & 2.269 & 1     & 0.125   & 1.75     & 2       &    & Exact  & \cite{PhysRev.65.117, PhysRev.85.808} \\ \hline
\end{tabular}
\caption{Critical temperatures and critical exponents for the Ising model on the Sierpinski carpet (SC) lattice and the Ising model on the square lattice, as reported in the literature. The boundary conditions (BC) can be either free (F) or periodic (P). The effective dimension is calculated using the Ising hyperscaling relation (equation (\ref{eq:hyperscaling1})).}
\label{tab:criticalexponents}
\end{table}

\section{Methodology}
\label{methodology}

With the help of previously developed algorithms, such as the Metropolis algorithm and the Swendsen-Wang algorithm, and by following Sornette's previously  published proposal \cite{Sornette1992CriticalPT, fraysse1993critical}, an algorithm was built that enables the Ising model, with its correspondence to the BCP model, to reach its critical state solely with self-contained information. This algorithm is based on a feedback loop that converts the standard unstable phase transitions of the Ising model and the percolation model into critical self-organized dynamics by adjusting the control parameter according to the order parameter depending also on the value of the correlation length. Subsequently, this algorithm was applied to three different lattices, each with the same number of spins ($N=4096$): the SC lattice with $k=4$, a square lattice with $L=64$ and a stochastic version of the Sierpinski carpet in which each of the 9 squares is removed with a probability of 1/9 at each iteration, also with $k=4$.\\

Based on what is described in Section \ref{SOCdynamics}, the main algorithm needs to create a connected path through the system to deliver heat to it. A connected path, equivalent to an infinite cluster, should touch two adjacent edges of the grid. So, if an infinite cluster exists, the temperature should increase. Otherwise, the temperature should decrease. To achieve this, clusters will be formed using the Swendsen-Wang algorithm \cite{swendsen1987nonuniversal}, which relies on the Coniglio bonds. The fundamental steps of this main algorithm are detailed in algorithm \ref{main-algorithm}.

\begin{algorithm}
\caption{Main Algorithm}\label{main-algorithm}
\begin{algorithmic}[1]
    \Statex /* \textbf{Step 1:} Initialize the temperature $T_i$ */
    \State $T \gets T_i$
    \For{$\text{iteration} \gets 1$ \textbf{ to } $\text{totalIterations}$}
        \Statex /* \textbf{Step 2:} Randomly select a state $\mu$ at the current temperature */
        \State \textsc{RandomlyChooseState}($\mu$, $T$)
        \Statex /* \textbf{Step 3:} Achieve equilibrium using Metropolis and Swendsen-Wang algorithms */
        \State \textsc{MetropolisAlgorithm}($\mu$, $T$)
        \State \textsc{SwendsenWangAlgorithm}($\mu$, $T$)
        \Statex /* \textbf{Step 4:} Save the cluster distribution from the Swendsen-Wang algorithm */
        \State $clustersDistribution \gets$ \textsc{GetClustersDistribution}()
        \Statex /* \textbf{Step 5:} Adjust the temperature based on cluster analysis */
        \If{\textsc{InfiniteClusterExists}()}
            \State $dT \gets$ \textsc{IncreaseTemperatureStep}()
            \State $T \gets T + dT$
        \Else
            \State $dT \gets$ \textsc{DecreaseTemperatureStep}()
            \State $T \gets T - dT$
        \EndIf
    \EndFor
\end{algorithmic}
\end{algorithm}

\section{Results}
\label{results}

\subsection{Algorithm performance}

The algorithm converges towards temperatures close to those reported in Table \ref{tab:criticalexponents}, in both cases, with fluctuations that decrease as the lattice size increases. Figure \ref{fig:numberofclusters}.A  illustrates the performance of the algorithm in the SC  when initiated in two distinct regimes: $T < T_c$ and $T > T_c$.  An effective approach to characterize the algorithm's behavior near the critical point involves examining the number of clusters normalized by the total number of spins in the lattice as a function of the distance to the critical temperature. Figure \ref{fig:numberofclusters}.A depicts the distribution of the number of clusters for  $T  < T_c$ and Figure \ref{fig:numberofclusters}.B for $T > T_c$. In the first regime, there is an exponential decay as the algorithm moves away from the critical temperature. In contrast, in the second regime, there is a linear increase as the algorithm moves away from the critical temperature. This difference in cluster behavior is intricately linked to the magnetic state of the model and the nature of correlations among clusters. In the ferromagnetic phase, observed for $T < T_c$, the model exhibits pronounced magnetic correlations attributed to the long-range order among magnetic moments. This results in the emergence of a single infinite cluster that increasingly dominates the space as the temperature continues to decrease. Conversely, in the paramagnetic phase, which occurs for $T > T_c$, the model is characterized by thermal noise that overpowers magnetic correlations. As the temperature increases, thermal fluctuations grow more influential, further promoting the fragmentation of clusters. This behavior signifies a transition to short-range order among magnetic moments. In \cite{Domb_1977, Domb_1975}, Domb \textit{et al.} identified a discontinuity in the average cluster size at the critical point, demonstrating results consistent with these findings and broadening the discussion of this behavior.\\

\begin{figure}[H]
\centering|
\includegraphics[scale=0.105]{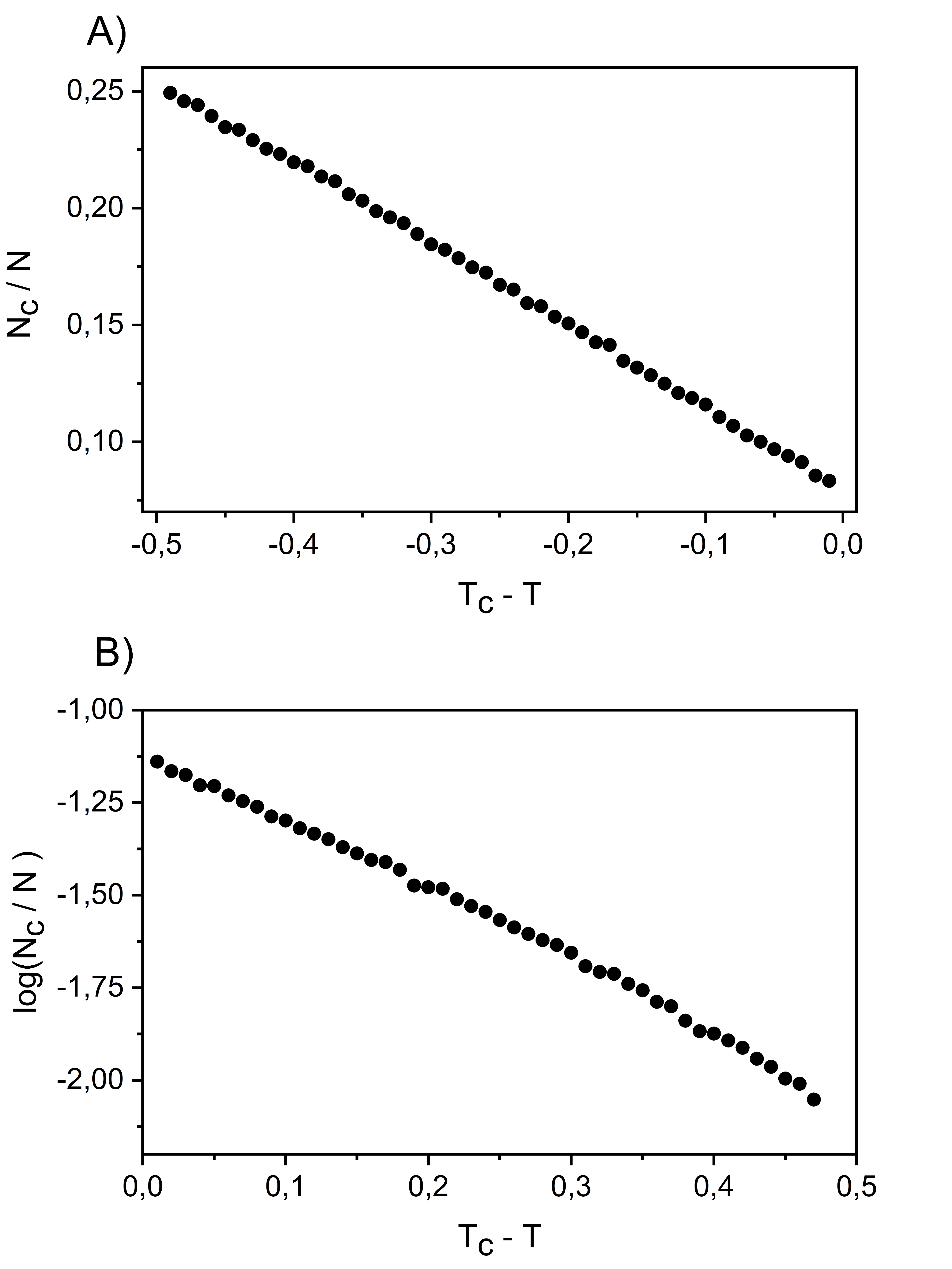}
\caption{\textbf{A)} Number of clusters evolution for $T_i < T_c$  in the SC lattice with $k=6$. This scenario shows an exponential decay in the number of clusters (semi-log plot). \textbf{B)} Number of clusters evolution for $T_i > T_c$ in the SC lattice with $k=6$. This scenario displays a linear increase in the number clusters. }
\label{fig:numberofclusters}
\end{figure}

Naturally, it is expected that the fractal structure had influenced the distribution of the clusters. This is because, unlike a compact model like a square, the formation of clusters is conditioned by the availability of spins, as the number of neighbors for each spin is variable and depends on its location. Figures \ref{fig:clusterdistribution-} and \ref{fig:numberofclusters+} show the behavior of clusters across three different types of lattices: the SC lattice, the stochastic version of the SC lattice and a square lattice.  Considering that all three lattices have the same number of spins ($N= 4096$), it is expected that as the number of clusters increases, the average cluster size decreases. Then, to quantify a difference in the distribution of clusters, the following quantity is introduced
\begin{equation}
    \eta = \frac{S_{mc}}{N_c}
    \label{eq:efficiency}
\end{equation}
where $S_{mc}$ is the mean size of the clusters and $N_c$ the number of clusters. It is noteworthy that these two quantities are interrelated. Specifically, $S_{mc} \cdot N_c = N$, where $N$ represents the total number of spins. Subsequently,
\begin{align}
     \eta = \frac{N}{N_{c}^{2}}
     \label{eq:eta_nc}
\end{align}
As $N$ is constant for the three cases, a large ratio indicates the presence of larger and more interconnected clusters. Meanwhile, a small ratio indicates that the clusters are smaller and more fragmented. In terms of assessing the algorithm's performance, a large $\eta$ ratio correlates with a higher efficiency in steering the system towards its critical state with minimal computational overhead. This efficiency stems from the reduced necessity to inspect and manage a myriad of smaller clusters. Large interconnected clusters simplify the system’s dynamics, making it easier and less computationally intensive to monitor and adjust the system’s parameters to reach and maintain criticality. Conversely, a small $\eta$ ratio, indicative of numerous small and fragmented clusters, demands greater computational resources for constant inspection and adjustment, thereby slowing down the process of achieving a critical state. This highlights the dual benefit of focusing on large cluster formations: they not only signal proximity to criticality but also enhance computational efficiency by simplifying the system’s management.\\

\begin{figure}[H]
\centering
\includegraphics[scale=0.31]{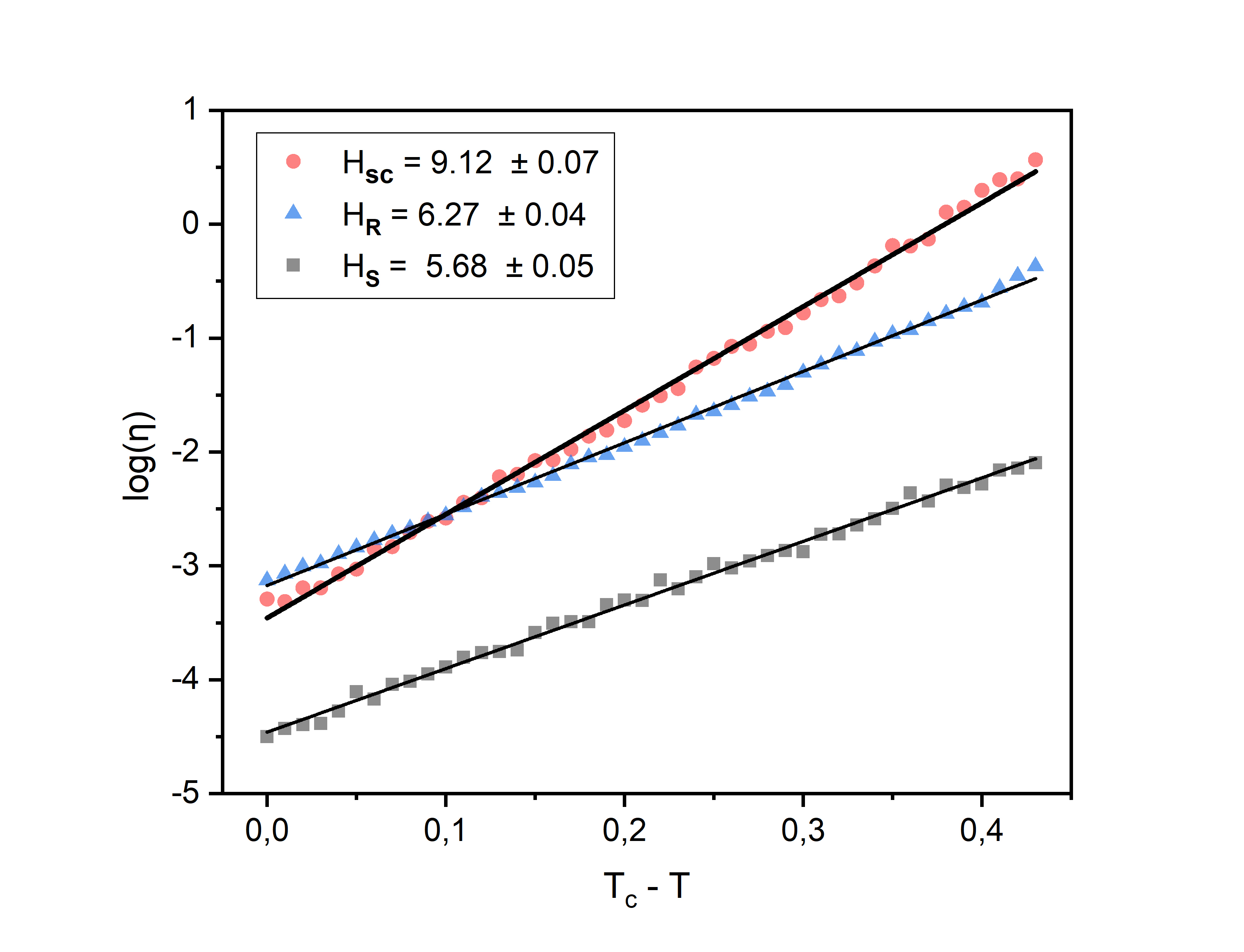}
\caption{Semilogarithmic plot showcasing the ratio between the mean cluster size and the number of clusters relative to the distance from the critical temperature for $T < T_c$, across three different lattice geometries: the SC lattice with $k=4$, a square lattice with $L=64$, and the stochastic version of the SC also with $k=4$ (all with $N=4096$). The slopes of these curves correlate with the exponential behavior of $\eta$, allowing for an intricate analysis of cluster interconnectivity and structure.}
\label{fig:clusterdistribution-}
\end{figure}

Figure \ref{fig:clusterdistribution-} illustrates the behavior of the quantity $\eta$ on a logarithmic scale, plotted as a function of the distance to the critical temperature, for temperatures $T$ less than the critical temperature $T_c$, in the three different lattice geometries.  It's crucial to underscore that for each lattice geometry, the algorithm identifies a unique critical temperature $T_c$ from which the distance to the critical point is measured. Each behavior is then quantitatively described by introducing the exponent $H$ as follows
\begin{align}
    \eta(T) \propto \exp \left(H(T_c - T)\right)
    \label{eq:eta}
\end{align}
For the SC, we find $H_{SC}= 9.12 \pm 0.07$. In contrast, the stochastic SC and the square lattice exhibit $H_{R}= 6.27 \pm 0.04$ and $H_{S}= 5.68 \pm 0.05$, respectively. These variations in $H$ values elucidate the relative interconnectivity of clusters and the facility with which each system gravitates towards a critical state. More specifically, an elevated $H$ value indicates a more rapid increase in the quantity $\eta$ as the temperature approaches the critical threshold $T_c$ from below, signifying more pronounced correlations or larger clusters as the phase transition is neared. This suggests a heightened tendency for the infinite cluster to fragment into a densely interconnected structure as it nears the percolation threshold, where it ultimately adopts a fractal dimension.\\

\begin{figure}[H]
\centering
\includegraphics[scale=0.31]{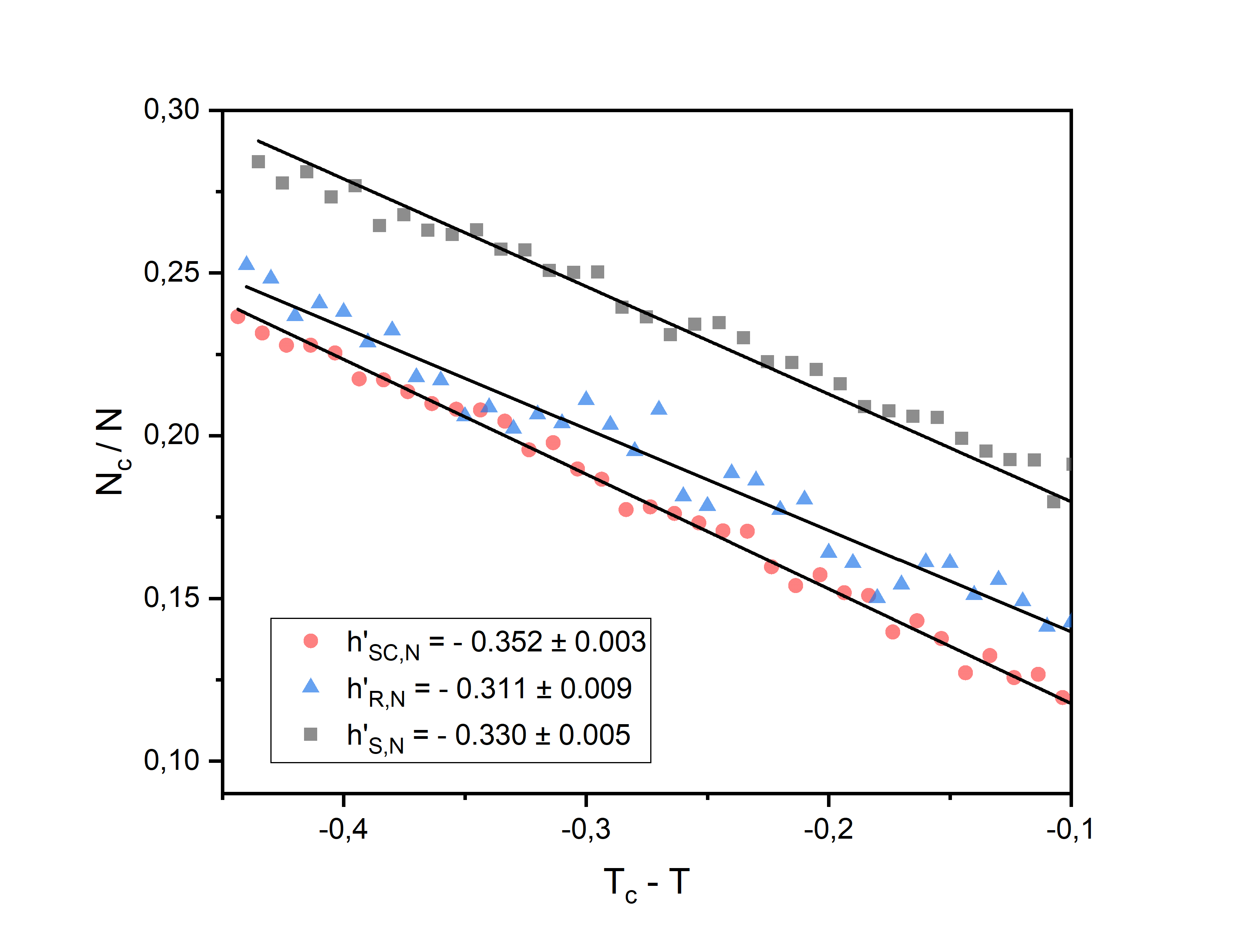}
\caption{Relationship between the number of clusters and the distance to the critical temperature for $T > T_c$, across three different lattice geometries: the SC lattice with $k=4$, a square lattice with $L=64$, and the stochastic version of the SC also with $k=4$ (all with $N=4096$). The graph illustrates a linear decrease in the number of clusters as the temperature increases beyond the critical threshold for each geometry. The slope of each line indicates the rate at which the number of clusters decreases with increasing temperature, shedding light on the distinct ways in which thermal fluctuations impact cluster dynamics across different structural configurations.}
\label{fig:numberofclusters+}
\end{figure}

Meanwhile, for temperatures $T$ greater than the critical temperature $T_c$, the number of clusters increases linearly with temperature. According to equation~(\ref{eq:eta_nc}), the quantity $\eta$ inversely depends on the square of the number of clusters. Then, in this regime, the relationship governing $\eta$ can be expressed as
\begin{align}
    \eta (T)= \frac{N}{(h^\prime(T-T_c)+N_{c_0})^2}
    \label{eq:eta2}
\end{align}
Here, $h^{\prime}$ represents the rate at which the number of clusters decreases as the temperature increases and $N_{c_0}$ accounts for the baseline number of clusters present at the critical temperature.\\

Our results show that the normalized values for  $h^\prime$ are $h'_{SC, N} = -0.352 \pm 0.003$, $h'_{R, N} = -0.311 \pm 0.009$, and $h'_{S, N} = -0.330 \pm 0.005$, as illustrated in Figure \ref{fig:numberofclusters+}. The steeper slope for the SC indicates that, as the temperature decreases, large clusters form more readily compared to the stochastic SC and the square lattices. However, in this regime, the differences are less significant than in the low-temperature regime because, at higher temperatures, clusters are smaller and less influenced by the lattice geometry. \\

For $T < T_c$, it is evident that the stochastic version of the SC exhibits slower convergence compared to the traditional SC but faster than that of a square lattice. This indicates that the stochasticity slightly hinders the convergence, but the system still gains advantages from the underlying fractal structure. Meanwhile, for $T > T_c$, the stochastic version of the SC shows slower convergence than both the traditional SC and the square lattice, primarily due to the randomness in its structure. At higher temperatures, where spins form much smaller clusters, this randomness hinders the convergence compared to the more compact square lattice. These findings suggest that the fast convergence observed in the classic version of the SC is primarily due to its hierarchical nature, which is diminished when randomness is introduced.\\

\begin{figure}[H]
\centering
\includegraphics[scale=0.105]{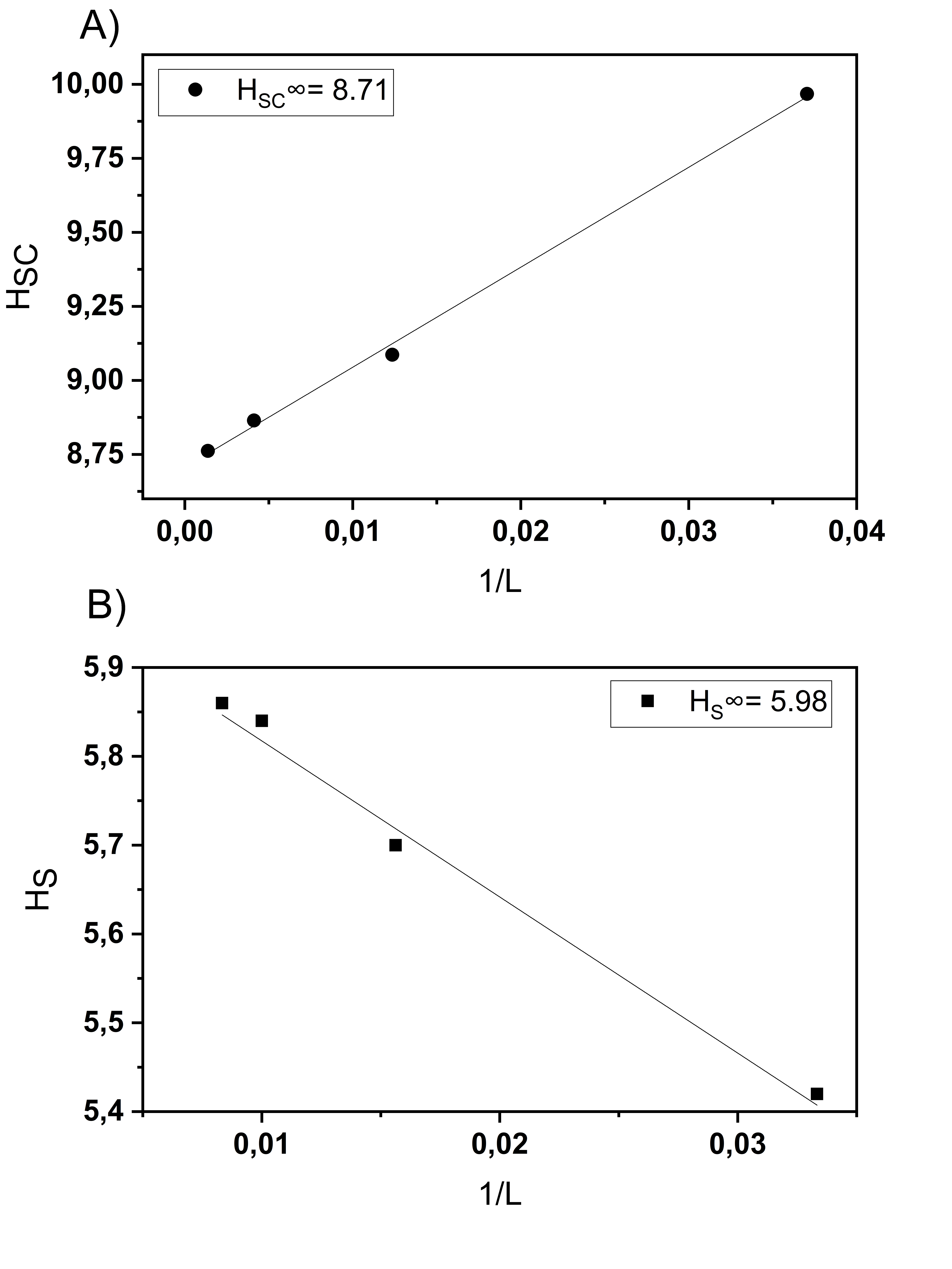}
\caption{Finite size scaling of the $H$ coefficient (equation (\ref{eq:eta})) in \textbf{A)} The SC lattice and \textbf{B)} The square lattice. $L$ represents the linear size of the lattice and the y-intercept in each plot represents the value of the $H$ coefficient in the thermodynamic limit, indicating the behavior as the lattice size approaches infinity.}
\label{fig:FSS}
\end{figure}

Finite-size scaling (FSS) analysis was conducted to investigate the behavior of the system as a function of lattice size for both the SC lattice and the square lattice. For the $H$ coefficients, the extrapolated values in the thermodynamic limit are presented in Figure \ref{fig:FSS}, yielding $H_{SC}^{\infty}= 8.71$ and $H_{S}^{\infty}= 5.98$. The results reveal distinct behaviors for each lattice type, which can be attributed to the different relationships between lattice size and the number of spins in each case. In the SC, as $L$ increases, the number of spins $N$ grows more slowly compared to the square lattice. Then, the negative finite-size correction arises because the compactness of the clusters becomes slightly less pronounced as the system size increases. In contrast, in the square lattice, clusters become relatively more compact as the system grows, leading to a positive correction. For the coefficient $h'$,  its variation across different lattice sizes was minimal, converging quickly to $h^{\prime \infty}_{SC, N} = - 0.35$ and  $h^{\prime \infty}_{S, N} = - 0.33$.

\subsection{Hyperscaling analysis and avalanche dynamics}

To enhance our understanding of the algorithm's performance on the SC, we numerically calculated the fractal dimension of the incipient infinite cluster $d_f$, describing its spatial scaling behavior; and the critical exponent $\tau$, detailing the size distribution of clusters at criticality.\\

\begin{figure}[H]
\centering
\includegraphics[scale=0.45]{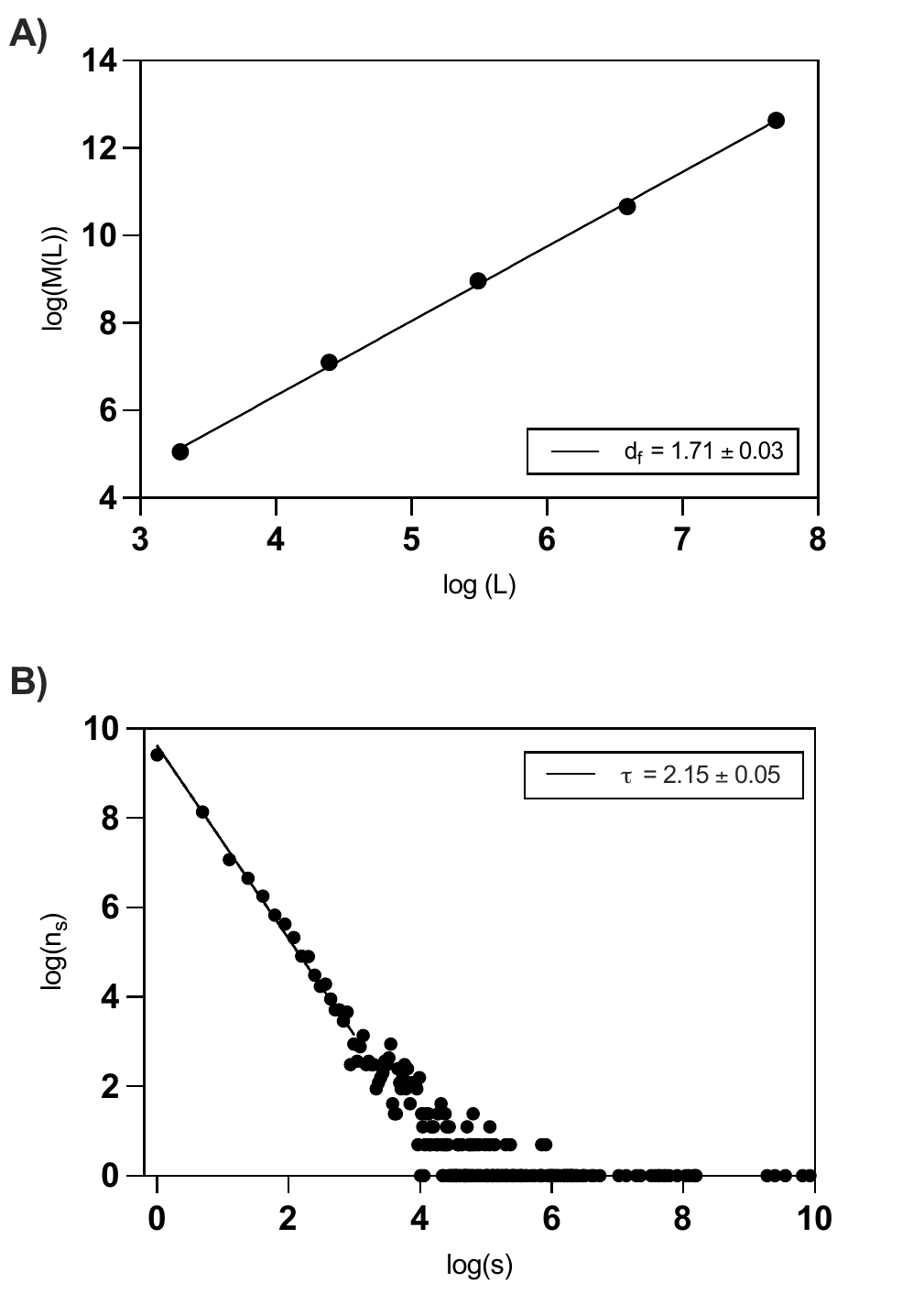}
\caption{\textbf{A.} Fractal dimension ($d_f$) of the incipient infinite cluster near the critical temperature for the Ising-BCP model on the Sierpinski carpet lattice. This estimation was done following equation (\ref{eq:fractaldim}). \textbf{B.} Cluster size distribution for the Ising-BCP model on the Sierpinski carpet lattice. This estimation was carried out using equation (\ref{eq:tau}), allowing for the calculation of the critical exponent $\tau$. It is noteworthy that the regression analysis was focused on the region where the decay is most pronounced. }
\label{fig:tau_and_df}
\end{figure}

Figure \ref{fig:tau_and_df}.A presents an estimation of the fractal dimension  of the incipient infinite cluster $d_f$ for the case of the SC. After the algorithm converged to the critical temperature, the mass of the infinite cluster, $M_*(L)$,  was calculated. Following equation (\ref{eq:fractaldim}), this process was repeated for various lattice sizes $L$. It is important to note that the increase in linear sizes actually corresponds to the increase in the iterations of the fractal construction ($k$), meaning that only lattices scaling as powers of 3 can be constructed. The graph clearly illustrates, through a logarithmic scale, a linear trend whose slope aligns with the sought-after value. Figure \ref{fig:tau_and_df}.B, on the other hand, presents the cluster size distribution for the Sierpinski carpet with $k=5$. To derive these values, at the critical point predicted by the algorithm, the frequency $n_s$ of each cluster size $s$ was computed. As specified by equation (\ref{eq:tau}), the slope value on a logarithmic scale determine the critical exponent $\tau$. The regression analysis was concentrated on the region where the decay is clearest.\\

Considering the critical exponents $\beta$ and $\nu$ specifically for free boundary conditions --since the concept of an infinite cluster is only meaningful in this context-- (see Table \ref{tab:criticalexponents}), and using equation (\ref{eq:fractal_beta}), we obtain  $D_{\text{eff}} = 1.86 \pm 0.03$. Conversely, using equation (\ref{eq:tau_df}), we obtain $D_{\text{eff}} = 1.97 \pm 0.09$. Both values are consistent with the fractal dimension of the Sierpinski carpet $D_f \approx 1.892$. This implies that the hyperscaling relations also hold for non-integer dimensions and that the mapping between the Ising and BCP model, upon which our algorithm is based, is effective in this fractal lattice. Table \ref{tab:lattice_properties} shows the fractal dimension of the percolating clusters and the $\tau$ exponent for different dimensionalities.\\

\begin{table}[H]
\centering
\begin{tabular}{|c|c|c|c|}
\hline
\textbf{Lattice dimension}          & \(\tau\) & \(d_f\) & \textbf{Reference}              \\ \hline
1.89 (SC) & $2.15 \pm 0.05$   & $1.71 \pm 0.03$  &  This work    \\ \hline
2 & $2.05 \pm 0.03$   & $1.90 \pm 0.06$ & \cite{cambier1986distribution}   \\ \hline
3  & $2.07 \pm 0.03$   & $2.30 \pm 0.05$ & \cite{cambier1986distribution}   \\ \hline
\end{tabular}
\caption{Fractal dimension of the percolating clusters and $\tau$ exponent for different dimensionalities.}
\label{tab:lattice_properties}
\end{table}

In addition, as detailed in Section \ref{SOCdynamics}, we initially described an avalanche as the flip of a single spin. However, since spins are interconnected according to Coniglio and Klein bonds, a more precise definition of an avalanche would be the simultaneous flip of all spins within a connected cluster. This process, where the entire cluster flips together based on the Metropolis rule, underpins the Swendsen-Wang algorithm, which is adept at simulating Ising dynamics and is employed in our simulations. Therefore, under this definition, the distribution of cluster sizes directly reflects the avalanche size distribution, which is depicted in Figure \ref{fig:tau_and_df}.B. It is clear that the distribution of avalanches in the system follows a power-law behavior, confirming that there is no characteristic scale for the avalanches, a hallmark of SOC. Furthermore, the fractal dimension of the avalanches can be inferred from the magnetic susceptibility exponent $\gamma$. This is because the number of spins influenced by the flip of a spin or a cluster scales as $L^{\gamma/\nu}$ in a system of size $L$. The values shown in Table \ref{tab:criticalexponents} thereby provide additional insight into the characterization of SOC behavior.

\section{Discussion}
\label{discussion}

For the SC, the simulations presented in Table \ref{tab:criticalexponents} have estimated the exponent $\nu$ to be within the range of $1.39$ to $1.73$ and $\gamma$ in the range of $2.22$ to $2.94$. On the other hand, the exact solution of the two-dimensional Ising model demonstrates that $\nu = 1$ and $\gamma = \frac{7}{4}$. For the correlation length, a higher exponent suggests that the  system's correlations extend over longer distances, meaning that fluctuations and interactions between distant parts of the system become more significant. For the susceptibility, a higher exponent indicates an increased system sensibility, suggesting an amplified response to external perturbations.\\

The outcomes of the algorithm corroborated the insights from the prior analysis. For the SC, both the low-temperature regime and the high-temperature regime exhibited enhanced connectivity in their cluster formation near the critical point. This indicates that the infinite cluster, through which this type of structure undergoes a phase transition, possesses a formation pattern that enables it to spread more efficiently across space. As a result, it underpins long-range correlations and heightened responsiveness to perturbations.\\

This phenomenon is hypothesized to occur because of the ordered and self-similar structure of the SC, where the voids or empty spaces within the lattice are uniformly distributed across various scales. It is expected that clusters may grow more quickly along the edges of those voids than they do in the interior of the lattice. Then, this even distribution of holes across the lattice in a geometrically regular manner creates a more conducive environment for the merging and growth of clusters during critical transitions. In other terms, the holes in the lattice create a guiding path that allows easier connections, as the clusters start growing at different scales and then easily find each other, promoting continued expansion.\\

In summary, our analysis reveals that the Ising-BCP model constructed on the SC lattice undergoes a phase transition characterized by increased connectivity within its infinite cluster. This connectivity, signified by a greater number of branches and connections, enhances the cluster's resilience to disruptions and facilitates a quicker response to perturbations. Moreover, it displays long-range correlations. These findings open up potential applications in information transmission, as systems can exploit long-range correlations to propagate information over extended distances. This can be used in various fields, including electrical models, neural networks, and transport networks. Additionally, another unexplored application is the one we developed during this paper: the construction of more efficient algorithms to drive systems toward SOC.

\section{Conclusion}

In conclusion, the research presented in this paper explored a self-organized critical dynamic on the SC lattice, shedding light on the emergent properties that this kind of structure exhibits. Building on previously reported proposals \cite{Sornette1992CriticalPT}, we conceptualized a dynamic capable of driving the Ising-BCP model towards SOC. Subsequently, we implemented a numerical algorithm embodying this dynamic. Additionally, for the first time, we determined the critical exponent $\tau$ and the fractal dimension $d_f$ of the SC's infinite cluster, with these findings being corroborated by hyperscaling relations.\\

When analyzing the critical exponents for the SC, we observed that, compared to a two-dimensional lattice, its phase transition exhibits greater long-range correlations and a more responsive adaptation to disturbances. The algorithm's effectiveness is mirrored in this behavior: large interconnected clusters not only promote the development of long-range correlations but also enhance the system's capability to quickly adapt to external perturbations, thereby ensuring robustness in its critical behavior. The adaptability revealed through the study of these clusters holds significant promise for technological advancements, such as the creation of more resilient algorithms and networks, as well as for gaining deeper scientific understanding of complex systems, including neural networks. \\

It is noteworthy that this enhanced efficiency is a consequence of the cluster formation on the SC: its iterative construction, ensuring the presence of holes at various scales, provides a guiding framework for clusters to form. However, it is crucial to acknowledge that these findings may have limited applicability to other fractal structures that haven't undergone a similar iterative process of systematically removing parts. Moreover, it is important to highlight that the computational simulations we have conducted have precision constraints influenced by the limited sampling of sizes inherent in the Sierpinski carpet geometry, as well as limitations in computational capacity for simulating very large lattices.\\

Despite the above, our study reveals an interesting dependence of self-organized critical dynamics on topology that had not been previously reported. In particular, the results of our simulations provide sufficient evidence that the SC, within and Ising-BCP model, achieves a SOC state more efficiently. We conclude that this is a consequence of its long-range correlations given its spatial distribution of spins and the impact this has on the formation of clusters. These results may have a variety of applications in any field that involves information transmission. 

\section*{Declaration of competing interest}

The authors declare that they have no known competing financial interests or personal relationships that could have appeared to influence the work reported in this paper.

\section*{Acknowledgments}

We thank Exacore HPC Uniandes for providing high performance computing time. GT acknowledges support from Fondo de Investigaciones, Facultad de Ciencias, Universidad de los Andes INV-2023-176-2951.

\bibliographystyle{unsrt}
\bibliography{references.bib}

\end{document}